\documentclass{eas}
\usepackage{graphicx}
\begin{document}

\title{The Elusive ISM of Dwarf Galaxies: Excess Submillimetre Emission \& CO-Dark Molecular Gas } 
\runningtitle{The Elusive ISM of Dwarf Galaxies}
\author{Suzanne C. Madden}\address{Service d'Astrophysique, CEA, Saclay, France} 
\author{Maud Galametz} \sameaddress{1} 
\author{Diane Cormier}\sameaddress{1} 
\author{Vianney Lebouteiller}\sameaddress{1} 
\author{Fr\'ed\'eric Galliano}\sameaddress{1} 
\author{Sacha Hony}\sameaddress{1}  
\author{Aur\'elie R\'emy}\sameaddress{1}  
\author{Marc Sauvage}\sameaddress{1} 
\author{Alessandra Contursi} \address{Max-Planck-Institute fuer extraterrestrische Physik, Garching, Germany}
\author{Eckhard Sturm}\sameaddress{2}
\author{Albrecht Poglitsch}\sameaddress{2}
\author{Michael Pohlen}\address{School of Physics and Astronomy, Cardiff University, Cardiff, Wales, UK}
\author{M.W.L. Smith}\sameaddress{3}
\author{George Bendo}\address{Astrophysics Group, Imperial College, London, UK}
\author{Brian O'Halloran}\sameaddress{4}
\author{The SPIRE SAG 2 and PACS consortia}

\begin{abstract}
  The Herschel Dwarf Galaxy Survey investigates the interplay of star formation
activity and the the metal-poor gas and dust of dwarf galaxies
using FIR and submillimetre imaging spectroscopic and photometric observations
in the 50 to 550 ~$\mu$m window of the Herschel Space Observatory. The dust SEDs are well constrained with the new Herschel and MIR Spitzer data. 
A submillimetre excess is often found in low metallicity galaxies, which,
if tracing very cold dust, would highlight large dust masses not easily
reconciled in some cases, given the low metallicities and expected gas-to-dust
mass ratios. The galaxies are also mapped in the FIR fine-structure lines (63
and 145~$\mu$m OI, 158~$\mu$m CII, 122 and 205 ~$\mu$m NII, 88~$\mu$m OIII) probing the low density ionised gas, the HII regions and photodissociation regions. While still
early in the mission we can already see, along with earlier studies, that line
ratios in the metal-poor ISM differ remarkably from those in the metal-rich
starburst environments. In dwarf galaxies, L[CII]/L(CO) ($\ge$ 10$^{4}$) is at least
an order of magnitude greater than in the most metal-rich starburst galaxies.
The enhanced [CII] arises from the larger photodissociation region where
 H$_{2}$, not traced by the CO, can exist. The 88~$\mu$m [OIII] line
usually dominates the FIR line emission over galaxy-wide scales in dwarf galaxies, not the 158 ~$\mu$m
[CII] line which is the dominant FIR cooling line in metal-rich galaxies. All
of the FIR lines together can contribute 1\% to 2\% of the L$_{TIR}$. The Herschel
Dwarf Galaxy survey will provide statistical information on the nature of the
dust and gas in low metallicity galaxies, elucidating the origin of the submm
excess in dwarf galaxies, and help determine a ([CII] +CO) to H$_{2}$ conversion
factor, thus providing observational constraints on chemical evolution models
of galaxies..
\end{abstract}
\maketitle
\section{Introduction}
The metal-enrichment of galactic interstellar medium (ISM) is a consequence of the life cycle of stars.  If we could map out the the metal-enrichment of galaxies throughout the history of the universe, we could unlock a key parameter controlling  the evolution of galaxies.  Present day telescopes can not easily open up the window to the low metallicity ISM at high-redshifts, but the local universe contains a vast zoo of accessible metal-poor laboratories to study the dust and gas properties under conditions that should resemble the early universe. Characterising the ISM  of dwarf galaxies has been challenging due to their low mass and relatively low infrared luminosities and the difficulty in assessing the molecular gas content. 
 
Previous infrared (IR) space missions,  such as ISO and Spitzer, as well as ground-based submillimetre (submm) and millimetre (mm) telescopes and the Kuiper Airborne Observatory, have noted the intriguing dust and gas properties of dwarf galaxies which set them apart from their metal-rich counterparts (e.g. Madden \cite{madden2000}; Galliano \etal\ \cite{galliano2003}; Galliano \etal\  \cite{galliano2005}; Madden \etal\ \cite{madden2006}; Wu \etal\  \cite{wu2007}; Engelbracht \etal\  \cite{engelbracht2008}; Galliano \etal\  \cite{galliano2008};  Leroy \etal\ \cite{leroy2009}; Galametz \etal\  \cite{galametz2009}; \cite{galametz2011}), notably, in these ways: 1) The abundance of PAHs relative to total dust mass is much lower 2) relatively warmer dust is often present over galaxy-wide scales, with a characteristically steeply rising mid-infrared (MIR) dust continuum and spectral energy distributions (SEDs) often peaking at shorter wavelengths - typically between 40 to 60 $\mu$m  3) excess emission at submm wavelengths is often seen $>$ 400 $\mu$m, which has been attributed in the past to large masses of cold dust  4) CO  detections are challenging, making it difficult to determine the molecular gas reservoir  5) ionic gas is prevalent, as seen from the high galaxy-wide [NeIII]/[NeII] line ratios. It has been difficult to be conclusive on these issues since the number of galaxies for which the full wavelength coverage of the dust and gas from the MIR to submm has been limited. However, this is improving now with the broad wavelength coverage, the spatial resolution and sensitivity of the Herschel Space Observatory (HSO; Pilbratt \etal\ \cite{pilbratt2010}). The Herschel Dwarf Galaxy Survey is a key program dedicated to surveying the gas and dust properties of dwarf galaxies of widely ranging luminosities, star formation activities and metallicities (Z).  Many targets of the Dwarf Galaxy Survey possess massive young star clusters and/or super star clusters (SSC), unusual modes of star formation with stellar surface densities orders of magnitude in excess of normal HII regions and OB associations.  
While Herschel observations are still underway, the initial observations, are already revealing the unique ISM of low metallicity galaxies.

\section{Dust Properties of Dwarf Galaxies - Excess Submillimetre Emission}
 
\begin{figure*}
\begin{center}
\includegraphics[width=6cm]{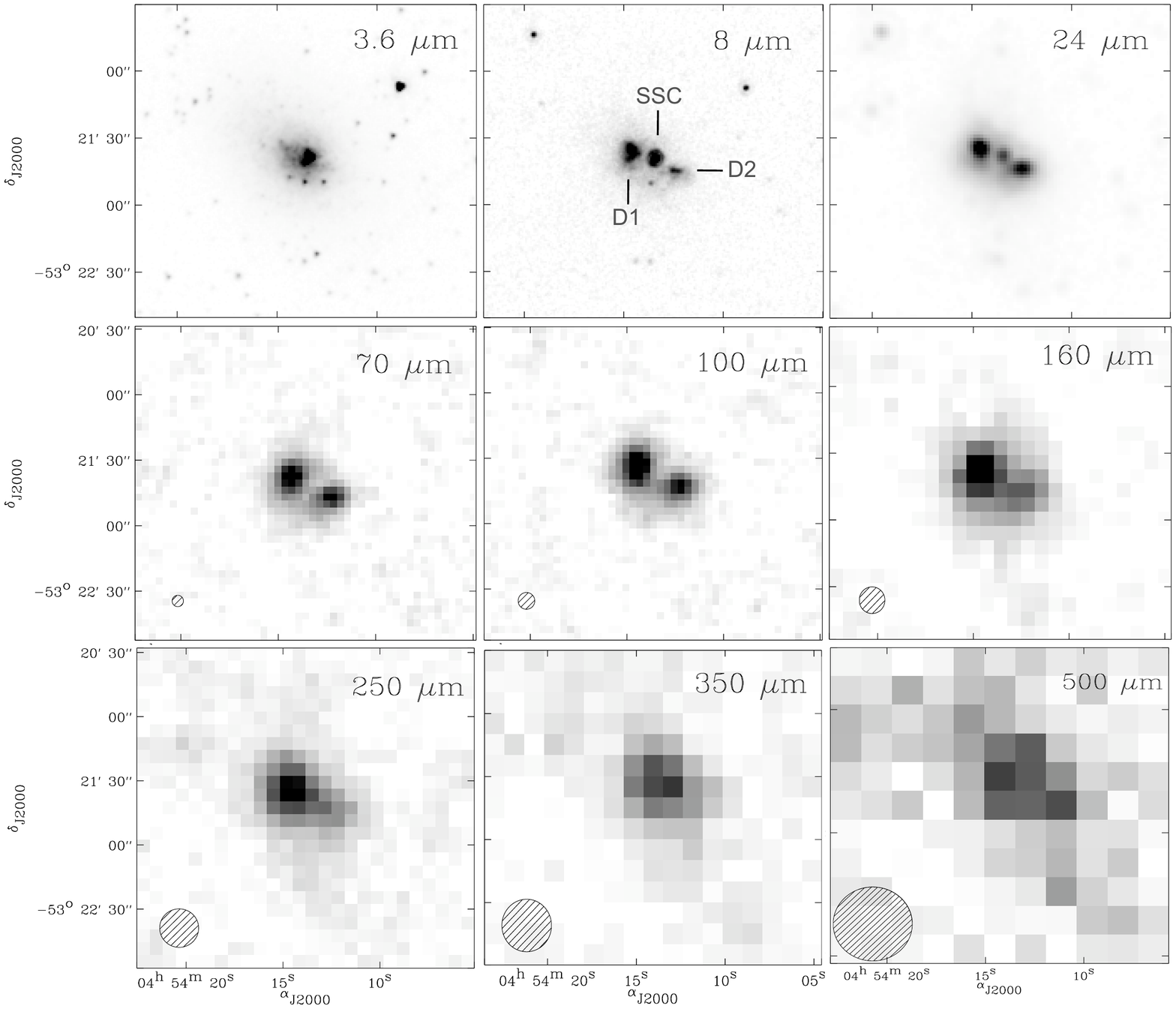}
\includegraphics[width=6cm]{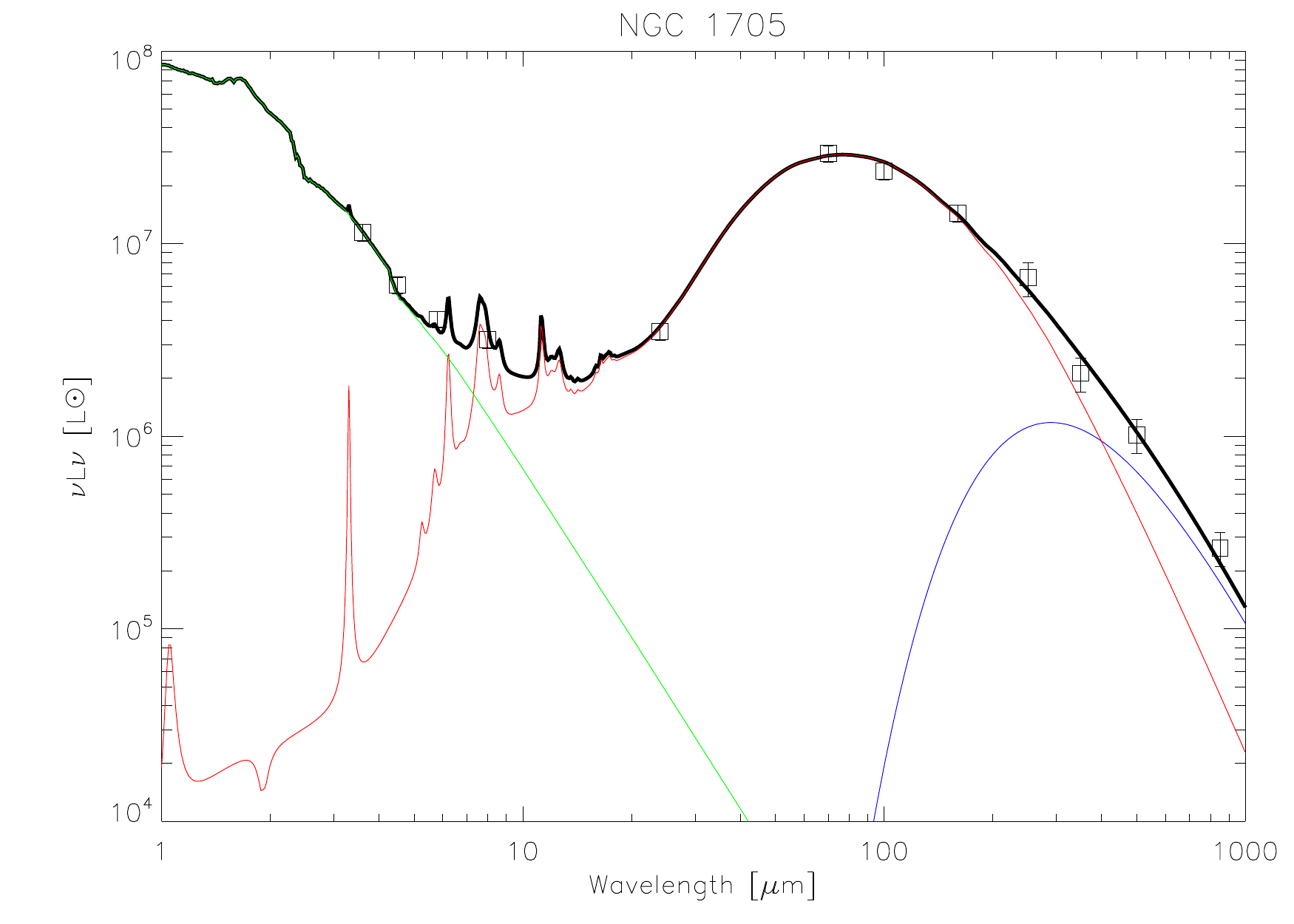}
\caption{(left) NGC~1705 viewed with Spitzer and Herschel. Note the super star cluster which is present only up to 24 $\mu$m.  Beam sizes of the Herschel instruments are represented as shaded circles. (right) The SED of NGC~1705 modeled with a stellar component (green) and dust emission (red). While several explanations exist for the excess submm emission, it is modeled here with a 10K component (blue). Spitzer and Herschel data points are the open boxes and the black line is the total modeled SED (O'Halloran \etal\ \cite{ohalloran2010})}.
\label{ngc1705figs}
\end{center}
\end{figure*}

NGC~1705 is a nearby (5 Mpc)  starbursting dwarf galaxy (Z= 0.35 Z$\odot$) completely dominated at UV and optical wavelengths by a super star cluster, the primary power source for the observed galactic outflow (Meurer \etal\  \cite{meurer2006}).  The remarkable variations in the 3.6 to 500 $\mu$m emission (Fig. \ref{ngc1705figs} )   illustrate the SSC fading beyond 24 $\mu$m, highlighting the preponderance of hot dust associated with the SSC, while  2 neighboring star forming sites, about 250 pc away on either side of the SSC, dominate the emission at FIR wavelengths and still further into submm wavelengths, the easternmost of these flanking infrared sources becomes predominant.  NGC~1705 is a valuable laboratory to explore the interplay between the very different star forming regions and the ISM.  It was mapped with PACS (70 to 160$\mu$m; Poglitsch \etal\ \cite{poglitsch2010}) and SPIRE (250 to 500$\mu$m; Griffin \etal\ \cite{griffin2010}) and combined with Spitzer MIR observations (Cannon \etal\ \cite{cannon2006}) and 870 $\mu$m Laboca/APEX data (Galametz et al 2009) to study the global 3.6 to 870 $\mu$m SED (O'Halloran \etal\  \cite{ohalloran2010}).  The new spatial information allows us to investigate the factors contributing to the different wavelength domains of the global SED.   Using the SED model of Galliano \etal\  (\cite{galliano2008}) which includes PAHs, silicates and graphite grains, O'Halloran \etal\  \cite{ohalloran2010} find excess submm emission, beyond $\sim$400 $\mu$m which can not be accounted for by the usual dust model. 
(Fig. \ref{ngc1705figs}).  

Meanwhile, several possibilities have been proposed for the presence of this submm excess emission in dwarf galaxies which has been noted before Herschel (e.g. Galliano \etal\ \cite{galliano2003}; Dumke \etal\ \cite{dumke2004}; Galliano \etal\  \cite{galliano2005}; Bendo \etal\ \cite{bendo2006}; B\"ottner \etal\ \cite{bottner2008}; Galametz \etal\  \cite{galametz2009}, \cite{galametz2011}), one being very cold dust ($\sim$10 K) and emissivity index ($\beta$) less than 2, which is normally used to explain the slope of the FIR to submm emission of galaxies (e.g. Draine \etal\ \cite{draine2007}). Using a  $\beta$ $\sim$~1 to account for a flatter submm slope, for example, as has been used in the past to explain the excess emsission, results in a large total dust mass for NGC~1705,  $1.8 \times10^5 M_{solar}$,  with 50\% of the dust mass residing at 10 K. The corresponding gas-to-dust  mass ratio (G/D) is  $\sim 100$,  lower than that expected from chemical evolution considering the low metallicity. Thus, either the total gas mass, observed to be mostly atomic (Meurer {\em et al.\/} \cite{meurer2006}), is underestimated, or the dust mass is overestimated. To test the effect of a different grain emissivity in the submm, for example, amorphous carbon was inserted in the model in lieu of graphite. Since amorphous carbon is more emissive than graphite, the resulting SED gives a lower dust mass, in this case, by about a factor of 4. The resulting G/D, using amorphous carbon, was found to be 435, which is the value expected for this low metallicity galaxy.  Thus a possible alternative explanation which alleviates the large dust masses found in the cold dust origin for the submm excess, is a more emissive dust component. Other explanations for the excess submm emission include spinning dust in ionised gas (e.g. Bot \etal\ \cite{bot2010};  Planck collaboration \cite{planck_collaboration}).

To zoom into the dust SEDs of individual star forming regions with more detail, the nearby (490 kpc) low metallicity galaxy, NGC~6822 was mapped with Herschel, providing the spatial resolution to model regions of a variety of activity (Galametz \etal\ \cite{galametz2010}). The distribution of the atomic gas and the dust at all wavelengths peak toward the prominent star forming regions, the most brilliant being Hubble V and Hubble X which contain very young stellar population of less than 10 Myr (Bianchi \etal\ \cite{bianchi2001}).  The dust SEDs of individual regions, as well as the total SED, in this case, do not show a submm excess, as was seen in NGC~1705,  which is also seen now in other low metallicity galaxies with Herschel. Large dust masses, however, are still found for NGC~6822 when using graphite as is originally used in the Galliano \etal\ (2008) model. Variations of the G/D over NGC~6822 range from 35 (Hubble V) to 80 (total galaxy SED) - extremely low for this galaxy of metallicity $\sim$ 30\% Z$_{solar}$.  This is, again, assuming that the majority of the gas mass is in atomic form and that molecular gas (Gratier \etal\ \cite{gratier2010}) accounts for a very small fraction of the total (less than 20\%). As in NGC~1705, the dust mass determined from the SED modelling can be decreased using amorphous carbon, with the consequence of decreasing the total G/D of NGC~6822 to $\sim$ 190, still lower than expected by more than a factor of 2. Thus the G/D ratios are very uncertain for dwarf galaxies.  
Herschel observations of the LMC (Meixner \etal\ \cite{meixner2010}) also show excess submm emission around 500 $\mu$m which would require uncomfortably large dust masses, if we are properly accounting for all of the gas mass with only the HI and little CO. Again, when amorphous carbon is used in lieu of graphite, the G/D mass ratio is  more in accord with the low metallicity.  Grossi \etal\  (\cite{grossi2010}) also see an excess in the submm emission in Virgo cluster dwarf galaxies which is difficult to explain.
Further Herschel observations of the Dwarf Galaxy Survey will shed light on this excess submm emission seen to date in dwarf galaxies. However, until we are more confident with the determination of the total gas mass reservoir, quantifying the total dust masses in dwarf galaxies may remain uncertain.

\section{The gas properties of dwarf galaxies -  CO-dark molecular gas }

Local galaxies as well as the high redshift universe rely on CO observations to determine the molecular hydrogen reservoir - a  necessary ingredient for forming stars. In most star forming galaxies, converting CO to molecular gas is assumed to be routine. Dwarf galaxies, however, while often rich in atomic hydrogen, are notoriously deficient in CO. Numerous surveys have tried hunting down the molecular gas content via CO measurements, which can often be a challenging endeavor (e.g. Leroy \etal\ \cite{leroy2009}), leading us to ask the question: where is the bulk of the fuel for the young star clusters and SSCs often present in dwarf galaxies?  
 
Herschel guaranteed time key programs are dedicated to quantifying the molecular gas content of low metallicity environments via their dust and gas measurements, using the FIR fine-structure lines as a means to probe the physical properties of the elusive molecular clouds.
The 158 $\mu$m [CII] line and the 63 and 145 $\mu$m [OI] lines are usually the most important coolants of the neutral gas, originating for the most part from the photodissociation regions (PDR) around molecular clouds and have been used to probe the radiation field, densities and temperatures. 

\begin{figure*}
\includegraphics[width=6.0cm]{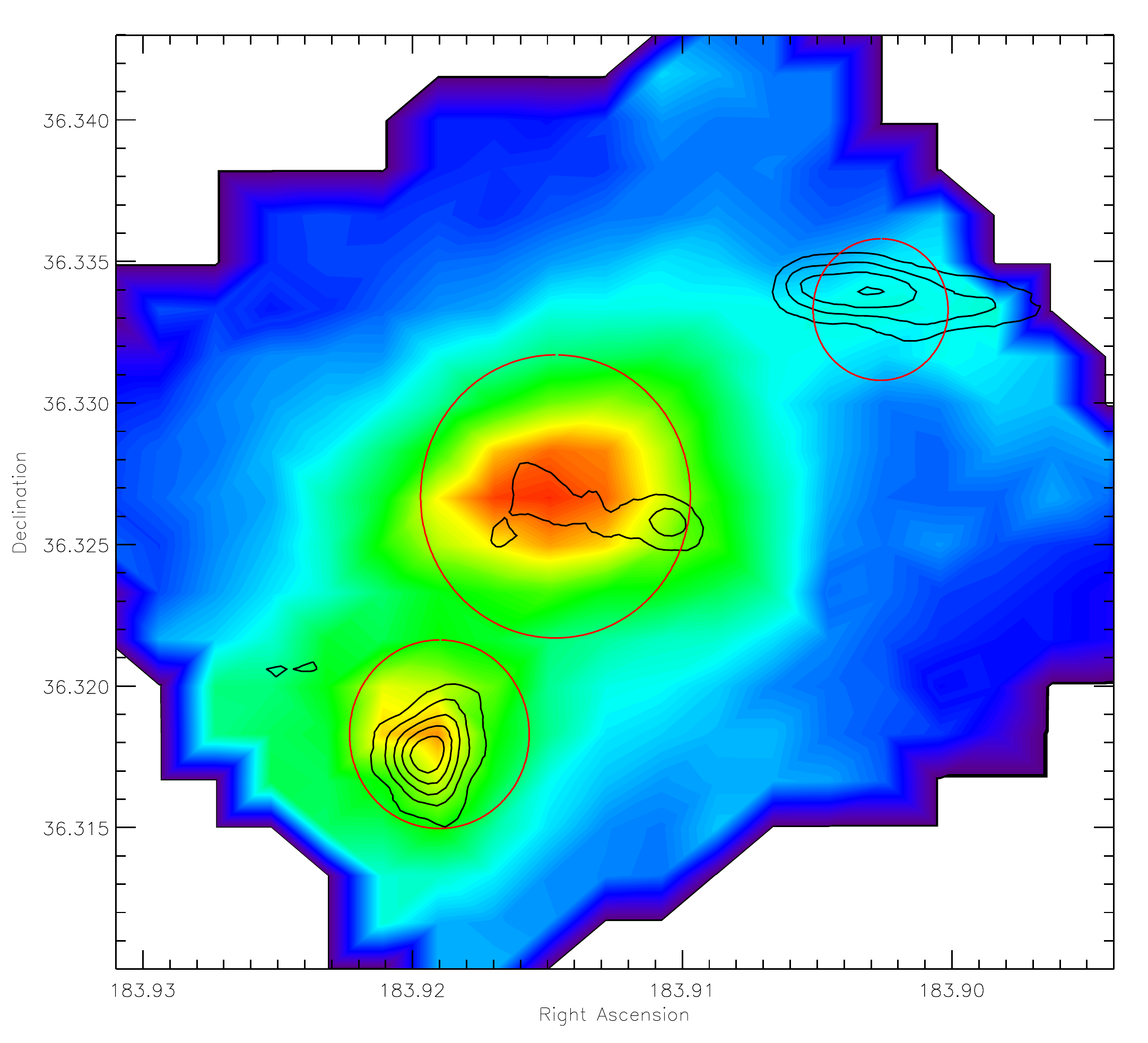}
\includegraphics[width=8.0cm]{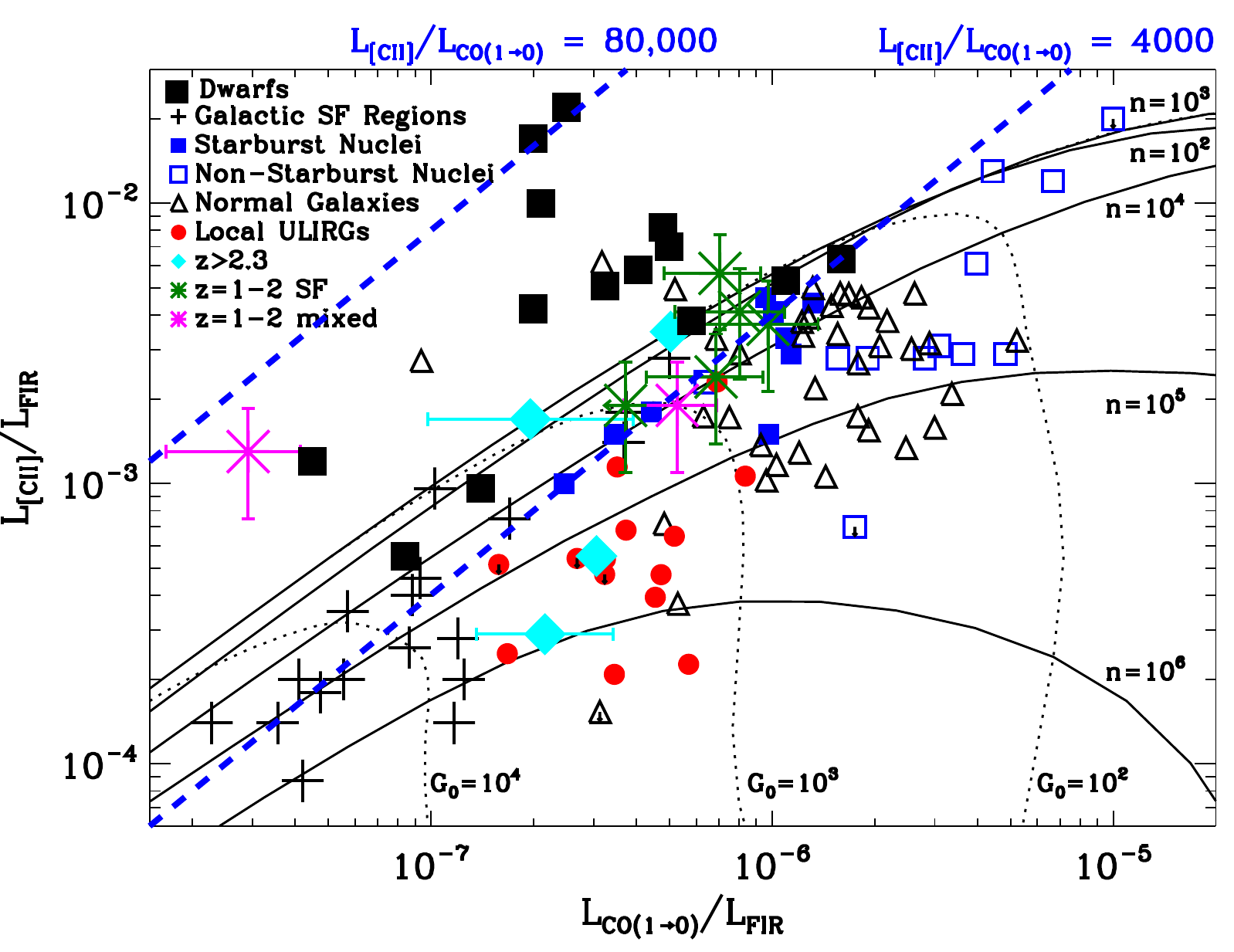}
\caption{(left) Image of 158 $\mu$m [CII] line emission in NGC~4214 with the CO contours (black) from Walter \etal\ (\cite{walter2001}). The red circles are the apertures used for PDR analysis in Cormier \etal\  (\cite{cormier2010}).  (right) L$_{[CII]}$/L$_{FIR}$ vs L$_{CO}$/L$_{FIR}$ for local galaxies, and high-z galaxies (Hailey-Dunsheath \etal\  \cite{hailey2010} and Stacey \etal\ \cite{stacey2010}) and low metallicity dwarf galaxies (\cite{madden2000}).  Lines of constant L$_{[CII]}$/L$_{CO}$ are shown (blue, dashed) for 4,000 (local starburst and high-z galaxies) and 80,000, the maximum end of dwarf galaxies. These observations are pre-Herschel data. Herschel observations of dwarf galaxies should populate this plot further to the upper left region.}
\label{ngc4214figs}
\end{figure*}

The nearby dwarf galaxy, NGC~4214 was mapped in 6 FIR fine structure lines probing the neutral and ionised gas (Cormier \etal\ \cite{cormier2010}). The 158 $\mu$m [CII] line, usually the most luminous of the FIR lines, originating from the photodissociation regions (PDRs) around molecular clouds, is very extended and shows 3 distinct peaks, 2 of them corresponding to prominent star forming regions (Fig \ref{ngc4214figs}). 
The L$_{[CII]}$/L$_{CO}$ of the 3 regions range from 4,700 (the northern, more quiescent region), to 75,000 in the central star forming region, where the SSC is located. Neither the [CII] nor the 63 $\mu$m [OI] lines, however, are the most brilliant, as is usually the case with dusty starburst galaxies. The 88 $\mu$m [OIII] line is 1.0 to 1.7 times more luminous than [CII] and peaks on the SSC, $\sim$ 80 pc  to the west of the [CII] peak. It requires 35 eV to ionise O$^{+}$ to O$^{++}$, yet, permeating galaxy-wide scales, such luminous [OIII] emission may be indicating that a large volume fraction of the ISM is subject to hard, ionising radiation, unlike the dusty starbursts. 

Earlier FIR spectroscopic observations of low-metallicity galaxies have noted the enhancement of 158 $\mu$m [CII] line emission compared to the observed CO  and the high L$_{[CII]}$/L$_{FIR}$ (e.g. Poglitsch \etal\ \cite{poglitsch1995}; Madden \etal\ \cite{madden1997}; Hunter \etal\ \cite{hunter2001}).   Dwarf galaxies can have observed L$_{[CII]}$/L$_{CO}$ values an order of magnitude or more larger than even the most active dusty starbursts in the local universe (Fig. \ref{ngc4214figs}; updated from Madden \cite{madden2000} using KAO and ISO data and from Hailey-Dunsheath \etal\  \cite{hailey2010} and Stacey \etal\ \cite{stacey2010}).  Such extreme L$_{[CII]}$/L$_{CO}$ values are the effects of the low metallicity and clumpy ISM:  a consequence of the lower dust abundance is a longer mean-free path of the FUV photons which penetrate molecular clouds, photodissociating the CO leaving a relatively larger C$^{+}$-emitting envelope surrounding a small CO core. Depending on certain conditions including the A$_{V}$ and radiation field, the self-shielding capability of H$_{2}$ can leave a potentially significant reservoir of H$_{2}$ in the C$^{+}$-emitting region (Roellig \etal\ \cite{roellig2006}; Wolfire \etal\  \cite{wolfire2010}; Shetty \etal\ \cite{shetty2010}).  This is the CO-dark molecular gas and the [CII] traces this "missing" molecular gas that is not probed by the CO. Assuming the [CII] is tracing this CO-dark H$_{2}$, {\it 10 to100 times more H$_{2}$ has been proposed to be 'hidden' in the C$^{+}$-emitting regions in dwarf galaxies compared to that inferred by CO} (e.g. Poglitsch \etal\ \cite{poglitsch1995}; Madden \etal\ \cite{madden1997}).  
 
 \section{Summary and prospectives}
The dust and gas properties in low metallicity galaxies appear to be different from their dusty counterparts. Often a submm excess is present and if it is due to cold dust, the total dust masses can be very large. The problem of low G/D masses, not in accord with the low metallicity, can be alleviated using more emissive grains, such as amorphous carbon. Remarkably large volumes of ionised gas can exist on galaxy-wide scales, a consequence of the lower dust abundance. Thus the low metallicity ISM probably contains small CO cores making up a small volume filling factor.The lower opacity conditions in dwarf galaxies make it favorable to use [CII] to probe the bulk of the molecular gas in low metallicity conditions, where CO is a poor tracer of molecular gas. We can determine more accurately the H$_{2}$ reservoir in low-metallicity galaxies via a {\it (CO + [CII]) to H$_{2}$ conversion factor}. 
Mapping the FIR lines in local dwarf galaxies will allow us to study the conditions governing the HI-H$_{2}$ interface under various low-metallicity conditions and quantify the molecular gas reservoirs more accurately. The 88 $\mu$m [OIII]  line may become the primary diagnostic for the high-redshift low-metallicity galaxies which will be targets for future submm observatories, such as ALMA. The Herschel Dwarf Galaxy survey will give us statistical information on the nature of the dust and gas in low metallicity galaxies promising to resolve some of the outstanding issues surrounding the ISM and star formation in low metallicity environments and thus provide constraints on the chemical evolution of galaxies.
 

\end{document}